\pdfoutput=1
\documentclass[conference]{IEEEtran}
\IEEEoverridecommandlockouts
\usepackage{cite}
\usepackage{amsmath,amssymb,amsfonts,lipsum}

\usepackage{algorithmic}
\usepackage{algorithm}
\usepackage{graphicx}
\usepackage{textcomp}
\usepackage{xcolor}
\usepackage{dsfont}
\usepackage{pgfplots}
\usetikzlibrary{spy}
\pgfplotsset{width=8.6cm,compat=1.13}
\pgfplotsset{
    every mark/.append style={solid}
}
\usepackage{cuted}
\def\BibTeX{{\rm B\kern-.05em{\sc i\kern-.025em b}\kern-.08em
    T\kern-.1667em\lower.7ex\hbox{E}\kern-.125emX}}
\begin{document}

\title{Transforming RIS-Assisted Passive Beamforming from Tedious to Simple: A Relaxation Algorithm for Rician Channel\\

}


\author{\IEEEauthorblockN{Xuehui Dong\textsuperscript{1}, Rujing Xiong\textsuperscript{1}, Tiebin Mi\textsuperscript{1}, Yuan Xie\textsuperscript{2}, Robert Caiming Qiu\textsuperscript{1}}
\IEEEauthorblockA{\textsuperscript{1} School of Electronic Information and Communications,Huazhong University of Science and Technology\\Wuhan 430074, China\\
\textsuperscript{2} Tandon school of engineering, New York university, NY, United States\\
Email:\{dong\_xh, rujing, mitiebin, caiming\}@hust.edu.cn, yx2732@nyu.edu}
}

\maketitle
\begin{abstract}
    This paper investigates the problem of maximizing the signal-to-noise ratio (SNR) in reconfigurable intelligent surface (RIS)-assisted MISO communication systems. The problem will be reformulated as a complex quadratic form problem with unit circle constraints. We proved that the SNR maximizing problem has a closed-form global optimal solution when it is a rank-one problem, whereas the former researchers regarded it as an optimization problem. Moreover, We propose a relaxation algorithm (RA) that relaxes the constraints to that of Rayleigh's quotient problem and then projects the solution back, where the SNR obtained by RA achieves much the same SNR as the upper bound but with significantly low time consumption. Then we asymptotically analyze its performance when the transmitter antennas $n_t$ and the number of units of RIS $N$ grow large together, with $N/n_t\rightarrow c$. Finally, our numerical simulations show that RA achieves over $98\%$ of the performance of the upper bound and takes below $1\%$ time consumption of manifold optimization (MO) and $0.1\%$ of semidefinite relaxation (SDR).
\end{abstract}


\section{Introduction}

Forward-looking technologies for the sixth generation (6G) communications have become a research hotspot in the wireless communication community. The most effective means to achieve ultra-high data rates is using new spectrum technologies to increase the wireless bandwidth and system capacity. The upward shift of the spectrum causes the weaker electromagnetic waves' diffraction and the faster-received power decay. Those will lead to a reduction in signal coverage and poor scatter signals. Among all technological works pertaining to 6G, RIS is one of the most eye-catching ideas and a promising technology \cite{dang2020should}\cite{9136592}.

The RIS, also called programmable metasurface, is a two-dimensional structure with many passive elements consisting of positive intrinsic-negative (PIN) diodes and microstrip lines. With real-time intelligent phase shifters, RISs enable dynamic control over the wireless propagation channel for passive beamforming\cite{9133142}. Especially if there is no line of sight (LoS) path between the base station (BS) and the user, the RIS could provide a reflected solid LoS beam compared to other scatter beams. We can manipulate this reflected LoS path channel by intentionally adjusting the shifters in RIS to improve the received power\cite{9551980}. 

However, it is a challenge to generate narrow beams in the specified direction efficiently. The difficulty lies in that the reflection coefficients of each array element of the RIS are the same, resulting in a complex non-convex quadratic problem with unit circle constraints as (\ref{first equation}) where $\mathbf{R}$ is a semi-definite matrix: 
\begin{equation}
    \begin{aligned}
            \max _{\mathbf{w}}&\ \mathbf{w}^{\dagger}\mathbf{R}\mathbf{w}
            \\
            \text{s.t.}&\ w_i\in\mathbb{C},\\&\ |w_i|=1,\forall i=1,\dots,N.
    \end{aligned}
    \label{first equation}
\end{equation}

Related works have proposed many good ideas on passive beamforming to overcome the above challenge. In\cite{8647620}, beamforming at the base station and the passive reflection coefficients at the RIS has been optimized using semidefinite relaxation (SDR). In \cite{8741198}, an alternating maximization algorithm has been proposed, with one adopting gradient descent for the RIS design while the other is a sequential fractional programming-based approach. One discrete beamforming algorithm has been proposed in \cite{9779545}, which approximates the global optimum with twelve quantization levels. Moreover, a data-driven deep reinforcement learning technique was proposed\cite{9110869}, and the manifold optimization (MO) methods have been introduced into the RIS passive beamforming problems in \cite{8855810}\cite{9405423}. The solution obtained by MO can be regarded as the global optimum because the problem (\ref{first equation}) turns out to be a convex problem in terms of Riemannian geometry\cite{AbsMahSep2008}.

However, it is still challenging to calculate the corresponding optimal beam in a very short coherence time, especially for large-scale RISs. In some communication scenarios where high mobility needs to be satisfied, we need to let the beam track the user in real-time to meet the service requirements due to the high directionality and low robustness of the RIS outcoming beam.

In this paper, we propose a simple but efficient algorithm with the idea of relaxation. Here are our contributions:
\begin{itemize}
    \item we prove that problem (\ref{first equation}) has the closed-form global optimal solution when $\mathbf{R}$ is rank-one in Theorem 1. 
    \item we propose that the receiving signal power remains constant as the phases of all units in the RIS change by the same phase difference in Remark 1.
    \item we provide a relaxation algorithm (RA) whose computational time consumption is below $1\%$ of that of the MO; meanwhile, the performance is close to that of MO whatever the $\text{rank}(\mathbf{R})$ is.
\end{itemize}
\section{System Model and Problem Evaluaion}
\begin{figure}[!t]
    \centering
        {\includegraphics[width=0.8\linewidth]{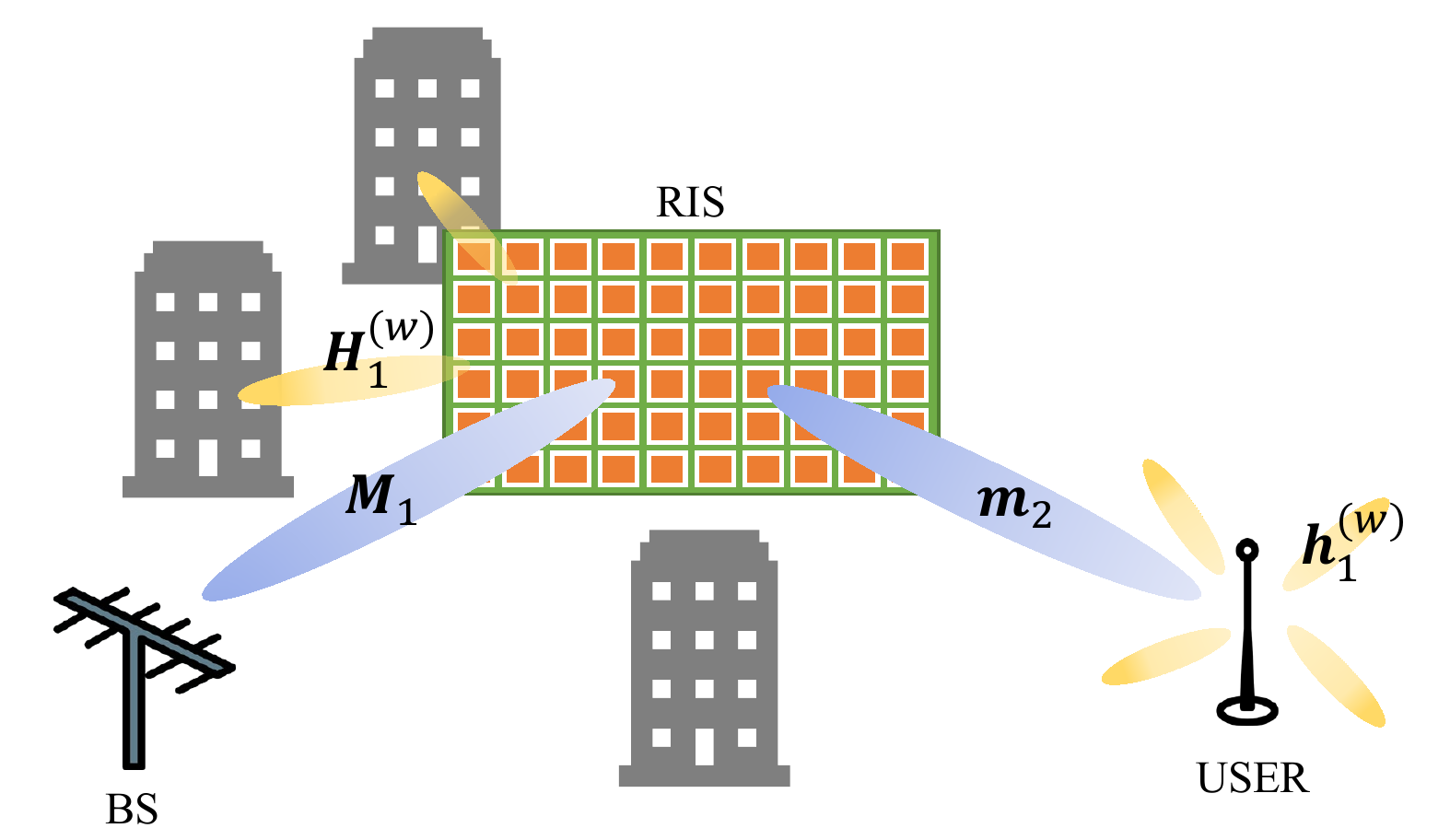}}
        \caption{A RIS-aided MISO system based on Rician channel model}
        \label{systems}
  \end{figure}

This section will establish the RIS-assisted reflected Rician channel model, which consists of the line-of-sight (LoS) path and the non-line-of-sight path between an $M$-antennas base station (BS) and one single-antenna user over a frequency flat fading channel as shown in Fig.\ref{systems}. Then we formulate the problem aiming to maximize the total transmit power as the expression of the problem (\ref{first equation}).
\subsection{System Model}
We assume that the system lacks a direct transmission path between the BS and the user here because it is blocked by big barriers, especially when the central frequency of the carrier wave is high. The Rician cascaded model relating the input signal vector $\mathbf{x}\in \mathbb{C}^{n_t}$ and output signal $y\in \mathbb{C}$ takes the form
\begin{equation}
    \begin{aligned}
        y=\mathbf{h}_2^{T}\mathbf{\Theta}\mathbf{H}_1\mathbf{x}+n,
    \end{aligned}
\end{equation}
where $[\cdot ]^T$ denoting the transpose, $\mathbf{H}_1\in\mathbb{C}^{N\times n_t}$ and $\mathbf{h}_2\in\mathbb{C}^{N\times 1}$ respectively denote the channel matrix from BS to RIS and from RIS, and $\mathbf{\Theta}=\textit{diag}(\alpha_1e^{j\theta_1},\dots,\alpha_Ne^{j\theta_N})\in\mathbb{C}^{N\times N}$ is a diagonal matrix representing the phase shift matrix of the RIS and $N$ is the number of units in the RIS. The noise scalar $n\in \mathbb{C}$ is a complex Gaussian scalar with zero mean and covariance $\sigma^2$.

For high-frequency communications, like millimeter waves or submillimeter waves, the channels are perhaps dominated by the LoS paths. To match the practical implementation, we employ the Rician fading to model the channel $\mathbf{H}_1$, which can be written as 
\begin{equation}
    \begin{aligned}
        \mathbf{H}_1=\sqrt{\frac{K_1}{K_1+1}}\mathbf{M}_1+\sqrt{\frac{1}{K_1+1}}\mathbf{H}_1^{(w)},
    \end{aligned}
    \label{5}
\end{equation}
where $\mathbf{H}_1^{(w)}$ is an i.i.d. matrix with zero mean, unit variance complex Gaussian entries, the rank-one matrix $\mathbf{M}_1$ is deterministic and arbitrary, normalized such that the channel gain $\text{tr}(\mathbf{M}_1\mathbf{M}_1^{\dagger})=n_t$ and $K_1$ is Rician factor between the two components. Different from the traditional communication systems, where the channel gain is usually related to both the number of transmitter's (Tx's) and receiver's (Rx's) antennas, the channel gain of $\mathbf{M}$ only depends on $n_t$ because of the lack of radio frequency (RF) chains in the RIS. Because the more elements RIS has, the more energy it reflects, then the channel gain of $\mathbf{h}_2$ depends on the number of the RIS's unit $N$ and Rx's antenna ($n_r=1$). So we also have 
\begin{equation}
    \mathbf{h}_2=\sqrt{\frac{K_2}{K_2+1}}\mathbf{m}_2+\sqrt{\frac{1}{K_2+1}}\mathbf{h}_2^{(w)},
\end{equation}
where $\mathbf{h}_2^{(w)}$ is an i.i.d. vector with zero mean, unit variance complex Gaussian entries, and $\mathbf{m}_2$ is a deterministic and arbitrary vector meanwhile $\text{tr}(\mathbf{m}_2\mathbf{m}_2^{\dagger})=N$ and the amplitude of each entry of $\mathbf{m}_2$ equals to $1$. Accordingly, the signal-noise-ratio (SNR) is given by 
\begin{equation}
    \text{SNR}=\frac{\Vert \mathbf{h}_2^{\dagger}\mathbf{\Theta}\mathbf{H}_1 \Vert^2 }{\sigma^2}.
    \label{SNR}
\end{equation}

\subsection{Problem formulation}
In this paper, our goal is to maximize the SNR at the receiver by optimizing the phase shift matrix $\mathbf{\Theta}$ with the application of maximum ratio transmission (MRT) at the RIS, subject to element-wise unit circle constraints\cite{8811733}. Accordingly, the problem can be formulated as
\begin{equation}
    \begin{aligned}
            \max _{\boldsymbol{\theta}}&\ \Vert \mathbf{h}_2^{T}\mathbf{\Theta}\mathbf{H}_1 \Vert^2 \\
            \text{s.t.}&\  0 \leq \theta_i \leq 2 \pi, \forall i=1,\dots,N,
    \end{aligned}
    \label{6}
\end{equation}
where $\boldsymbol{\theta}=[\theta_1,\dots,\theta_N]^T$, and the operator $\left\|\cdot\right\|$ is Frobenius norm. We assume that $\alpha_i=1,\forall i=1,\dots, N$ because the reflected coefficients of all units in RIS are the same; thus we let phase shift vector $\boldsymbol{w}=[e^{\theta_1},\dots,e^{\theta_N}]^H$. Next, by $\mathbf{h}_2^{T}\mathbf{\Theta}\mathbf{H}_1=\boldsymbol{w}^{\dagger}\boldsymbol{\Phi}$, where $\boldsymbol{\Phi}=\text{diag}(\mathbf{h}_2^{T})\mathbf{H}_1\in \mathbb{C}^{N\times n_t}$, we have
\begin{equation}
    \begin{aligned}
            \max _{\boldsymbol{w}}&\ \boldsymbol{w}^{\dagger}\mathbf{R}\boldsymbol{w} \\
            \text{s.t.}&\ |w_i|=1,\forall i=1,\dots,N,
    \end{aligned}
    \label{aaa}
\end{equation}
where $\mathbf{R}=\boldsymbol{\Phi}\boldsymbol{\Phi}^{\dagger}$ is a semi-definite matrix. So we can find that problem (\ref{aaa}) is exactly a complex non-convex quadratic problem with unit circle constraints as the problem (\ref{first equation}). 
\section{Relaxation Algorithm and Intuition Interpretation}
\subsection{Problem statment}
Since the constraint $|w_i|=1, i=1,2,\dots,n$ is the product space of n unit circles $\mathbb{S}^1$, which is also called n-dimensional torus $\mathbb{T}^N=\mathbb{S}^1\times\cdots\times\mathbb{S}^1$ in geometry. From the perspective of manifold, we take an element-wise mapping $\varphi$ on $\mathbf{w}$ from $\mathbb{T}^n$ to $(\mathbb{R}/2\pi\mathbb{Z})^N$, where the $\varphi: w_i\in\mathbb{C}\to \theta_i\in\mathbb{R}/2\pi\mathbb{Z},\forall i=1,\dots, N$. So the problem can be rewritten as 
\begin{equation}
\begin{aligned}
    \max_{\boldsymbol{\theta}\in \mathbb{R}^n}\ &\boldsymbol{w}^{\dagger}\mathbf{R}\boldsymbol{w}.\\
    \label{10}
\end{aligned}
\end{equation}
We perform an eigenvalue decomposition of the complex hermitian matrix $\mathbf{R}$. Assume that the rank of $\mathbf{R}$ is $n_t$, and that
\begin{equation}
    \mathbf{R}=\sum_{i=1}^{M}\lambda_i \boldsymbol{v_i}\boldsymbol{v_i}^{\dagger},
    \label{11}
\end{equation}
where $\lambda_i$ denotes $i^{th}$ eigenvalue in descending order (i.e. $\lambda_1\geq\lambda_2\dots\geq\lambda_M$), $M$ denotes the $\text{rank}(\mathbf{R})$ and $\boldsymbol{v_i}$ denotes the corresponding eigenvector. And we let $\boldsymbol{v}_i=\left[a_{i,1}e^{j\tau_{i,1}},\dots,a_{i,N}e^{j\tau_{i,N}}\right]^T$ which satisfy $\sqrt{a_{i,1}^2+a_{i,2}^2+\dots+a_{i,N}^2}=1$.

\subsubsection{$M=1$}
Here we proposed a significant theorem so that all the rank-one unit circle constrained complex quadratic problems have no need to be optimized by traditional approaches. All iterative algorithms for solving this kind of rank-one problem are meaningless.

\textit{\textbf{Theorem 1}: Problem (\ref{10}) has the closed-form global optimal solution when $\mathbf{R}$ is rank-one, where the optimal solution is
\begin{equation}
    \theta_i=\tau_{1,i}+C,\ \forall i=1,\dots,N,
    \label{remark one}
\end{equation}
where $C$ is a constant for all $\theta_i$ and $\tau_{1,i}$ is the phase of the $i^{\text{th}}$ entry of eigenvector $\boldsymbol{v_1}$.
}

\textit{\textbf{Proof}:
We consider the self-adjoint property of the hermitian matrix, and Euler's equation, Eq.(\ref{10}) can be derived as follow:
\begin{equation}
    \begin{aligned}
        \boldsymbol{w}^{\dagger}\mathbf{R}\boldsymbol{w}=&\lambda_1\boldsymbol{w}^{\dagger}\boldsymbol{v_1}\boldsymbol{v_1}^{\dagger}\boldsymbol{w}\\=&\lambda_1\sum_{i=1}^{N}\sum_{k=1}^{N} a_ie^{j(\tau_{1,i}-\theta_i)}a_ke^{-j(\tau_{1,k}-\theta_k)}.
    \end{aligned}
    \label{M=1}
\end{equation}
By the Hermitian property of Eq.(\ref{M=1}), we obtain $\boldsymbol{w}^{\dagger}\mathbf{R}\boldsymbol{w}$ equals to its real part $\Re\{\boldsymbol{w}^{\dagger}\mathbf{R}\boldsymbol{w}\}$  so that it can attain its maximum value by letting the phase of each part of it be zero, i.e.,
\begin{equation}
    \tau_{1,i}-\theta_i=\tau_{1,j}-\theta_j,\forall i,j=1,\dots,N.
    \label{close-form}
\end{equation}
}

\textit{\textbf{Remark 1}: The value of the objects function of Problem (\ref{10}) remains unchanged when $\theta_i, \forall i=1,\dots,N$ increase the same skewing $\Delta\theta$ (i.e., $\theta_i+\Delta\theta, \forall i=1,\dots,N$) whatever the rank of $\mathbf{R}$ is.
}

The conclusion above results from the following equation:
\begin{equation}
    \begin{aligned}
        \mathbf{w}^{\dagger} \mathbf{R}\mathbf{w}=
        \sum_{k=1}^{n-1}|r_{kk}|^2+\sum_{i=1}^{n-1}\sum_{j=i+1}^{n}2|r_{ij}|\cos{(\theta_i-\theta_j-\phi_{ij})},
        \label{bbb}
    \end{aligned}
\end{equation}
where $r_{ij}$ and $\phi_{ij}$,$\forall i,j$ are the amplitude and the phase of the entry at $i^{th}$ row and $j^{th}$ column of $\mathbf{R}$. Eq.(\ref{bbb}) illustrates the fact that what is meaningful in passive beamforming is the phase differences. 

\begin{figure}[!t]
    \centering
        {\includegraphics[width=0.7\linewidth]{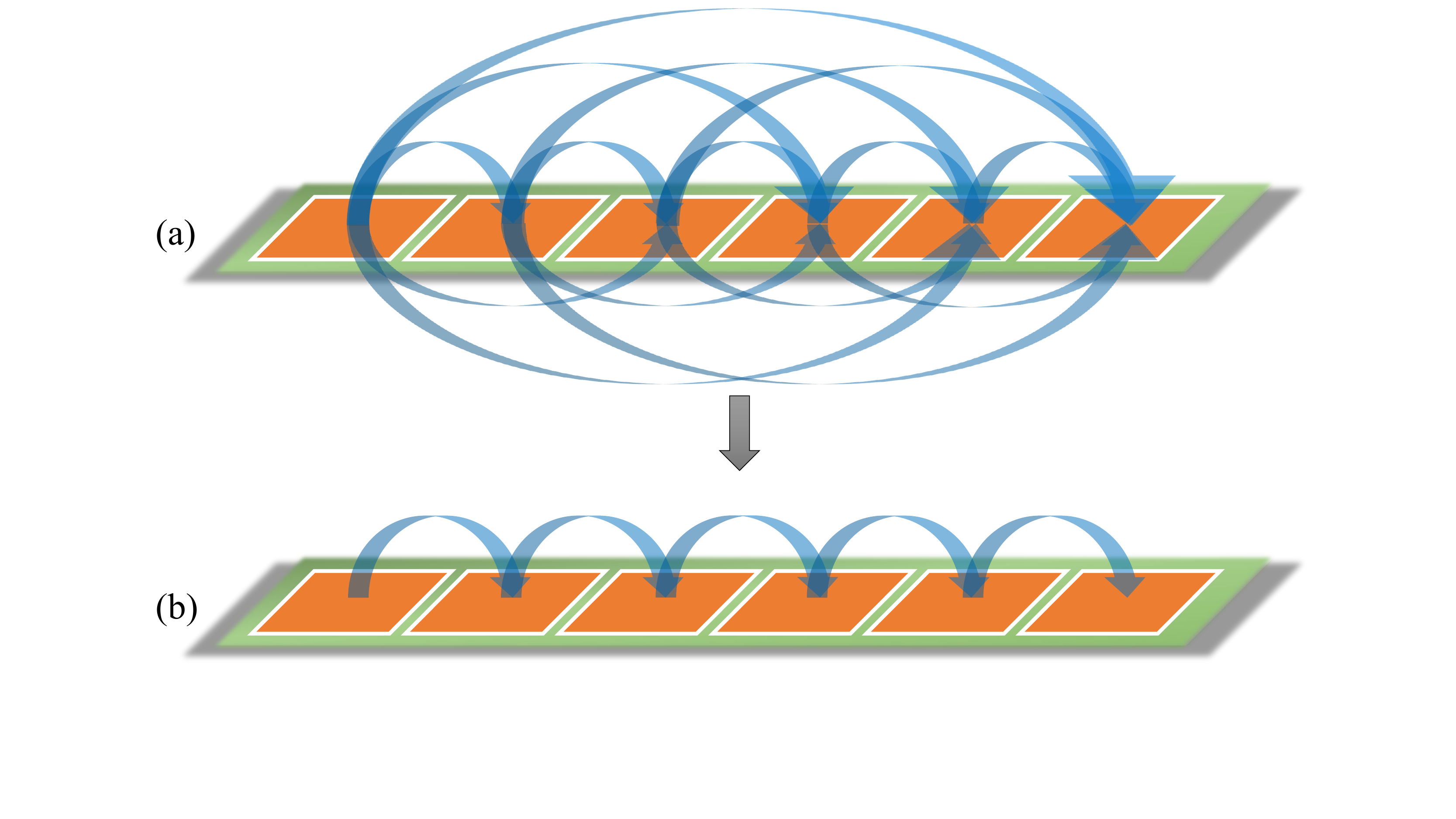}}
        \caption{An example of 6 units linear RIS shows that the optimization process transforms from tedious to simple, where the curved arrows indicate the phase differences. (a) all the 15 phase differences between the state of each unit with redundant quantities; (b) the five intrinsic phase differences without redundance}
        \label{decoupling}
\end{figure}

As shown in Fig.\ref{decoupling}, by Remark 1, we can intuitively simplify the optimization process of beamforming when there is only one planar incoming electromagnetic (EM) wave.
Constructive or destructive interference depends on the relative phase differences between each beam at the receiver. What matters is the phase difference between each unit (i.e., $\theta_i-\theta_j,\forall i,j$) as shown in Fig.\ref{decoupling}(a) and (\ref{bbb}). Actually there are only $N-1$ non-correlated phase differences among $\frac{N(N-1)}{2}$ . So we neglect the redundance ones and only consider the intrinsic ones, just like decoupling shown as fig.\ref{decoupling}(b).

\subsubsection{$M \geq 1$}
We can decompose the rank-$M$ problem (\ref{10}) to the linear combination of $M$ rank-one problem (\ref{10}) :
\begin{equation}
    \begin{aligned}
        \boldsymbol{w}^{\dagger}\mathbf{R}\boldsymbol{w}=\sum_{i=1}^{M}\lambda_i\boldsymbol{w}^{\dagger}\boldsymbol{v_i}\boldsymbol{v_i}^{\dagger}\boldsymbol{w}.
    \end{aligned}
    \label{14}
\end{equation}

For convenience, we denote the phases vector of the eigenvector $\boldsymbol{v_i}$ as $\boldsymbol{\tau_i}$ and $\tau_{i,j}$ represents the $j^{th}$ elements of $\boldsymbol{\tau_i}$. Ideally, we hope to get the maximum value of Eq.(\ref{14}) by Theorem 1. Unfortunately, there does not exist a set of constants $C_i$ such that for all $i,j=1,\dots,N$ the $\boldsymbol{\tau_i}+C_i=\boldsymbol{\tau_j}+C_j$ because of the orthogonality of eigenvectors set $\{\boldsymbol{v_i}\}_i^N$. 

So that our proposed algorithm obtains the approximated global optimum by simply applying the closed-form optimal solution (\ref{remark one}) into the first component $\lambda_1 \mathbf{w}^{\dagger}\boldsymbol{v_1}\boldsymbol{v_1}^{\dagger}\mathbf{w}$ of Eq.(\ref{14}), the margin of error between the result solved by RA and the global optimum is caused by the projection of $\mathbf{w}$ on other eigenvectors $\boldsymbol{v_i},i\neq 1$, which we are going to give the mathematical explanation after we state our relaxation algorithm.

\subsection{Algorithm description}
The idea of our relaxation algorithm is only dealing with the leading component $\lambda_1\boldsymbol{w}^{\dagger}\boldsymbol{v_1}\boldsymbol{v_1}^{\dagger}\boldsymbol{w}$ of $\boldsymbol{w}^{\dagger}\mathbf{R}\boldsymbol{w}^{\dagger}$ while ignoring the others $\sum^{M}_{i=2}\lambda_i\boldsymbol{w}^{\dagger}\boldsymbol{v_i}\boldsymbol{v_i}^{\dagger}\boldsymbol{w}$.

\begin{algorithm} 
    \caption{Relaxation Algorithm (RA)} 
    \label{relaxation} 
    \begin{algorithmic}[1] 
    \REQUIRE complex vector of variables $\boldsymbol{w}\in\mathbb{C}^{N\times 1}$, complex quadratic form coefficient matrix $\mathbf{R}\in\mathbb{C}^{N\times N}$ 
    \ENSURE optimal vector $\boldsymbol{w}^*$ 
    \STATE perform spectral decomposition of matrix $\mathbf{R}$ as (\ref{11}) and obtain the leading eigenvectors $\boldsymbol{v_1}$ 
    \STATE make $\boldsymbol{\theta}^*$ equals to the phase vector $\boldsymbol{\tau_1}$ of $\boldsymbol{v_1}$
    \STATE obtain the optimal solution vector by $\boldsymbol{w}^*=\boldsymbol{e}^{j\boldsymbol{\theta}^*}$ 
    \end{algorithmic} 
\end{algorithm}

Step 2 of Algorithm 1, which only considers the leading component corresponding to the largest eigenvalue, transforms the process of searching for optimal solution in the N-dimensional torus $\mathbb{T}^N$ from the tedious well-designed algorithm to a simple spectrum decomposition of complex hermitian matrix $\mathbf{R}\in\mathbb{C}^{N\times N}$. Apparently, the RA can directly obtain the maximum value of problem (\ref{10}) while $n_t=1$ as mentioned by Theorem 1. 

Let us explain the RA from the perspective of semi-positive definite relaxation\cite{5447068}. For the Rayleigh's quotient (RQ), which is constrained by $\boldsymbol{w}^{\dagger}\boldsymbol{w}=N$ ($C$ is an arbitrary constant), the objects function $\boldsymbol{w}^{\dagger}\mathbf{R}\boldsymbol{w}$ attains its maximum value $N\lambda_1$ (leading eigenvalue) when $\boldsymbol{w}=\boldsymbol{w}^*_{RQ}=\sqrt{N}\boldsymbol{v_1}$ (leading eigenvector). That is because $\boldsymbol{w}^*_{RQ}$ is not projected onto any other eigenvector. Noticeably, the set of complex vectors in n-dimensional torus $\mathbb{T}^N$ is a subset of complex vectors in the RQ problem's constraint space $\mathbb{S}^N$, which means
\begin{equation}
    \{\boldsymbol{w}\in\mathbb{C}^N| |w_i|=1,\forall i=1,\dots,N\}\subset \{\boldsymbol{w}\in\mathbb{C}^N|\boldsymbol{w}^{\dagger}\boldsymbol{w}=N \}.
    \label{affiliation}
\end{equation}

The geometry explanation of the relaxation process in the RA can be described as follow: 
\begin{enumerate}
    \item relaxing the feasible set of problem (\ref{10}) from $\mathbb{T}^N$ to $\mathbb{S}^N$ meanwhiles the problem converts into an RQ problem;
    \item obtaining the solution of RQ problem $\boldsymbol{w}^*_{RQ}$;
    \item projecting  the optimal solution $\boldsymbol{w}^*_{RQ}$ from  $\mathbb{S}^N$ to $\mathbb{T}^N$ by making amplitude of each entry of $\boldsymbol{w}^*_{RQ}$ equal to $1$, then obtaining the solution of the RA, i.e., $\boldsymbol{w}^*$.
\end{enumerate}

\subsection{Spectral Analysis}
\begin{figure}[!t]
    \centering
        {\includegraphics[width=0.7\linewidth]{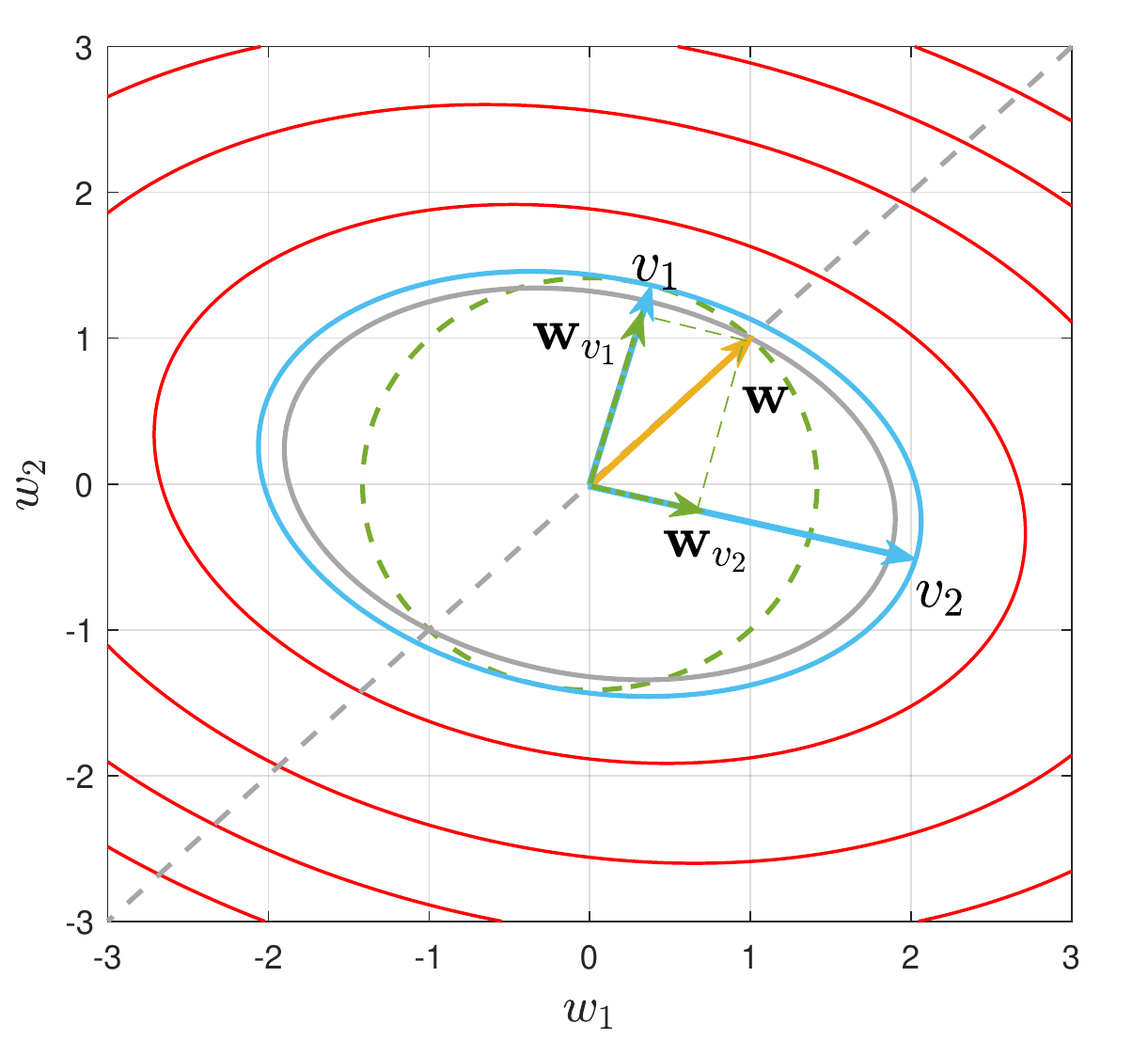}}
        \caption{A Rayleigh's quotient problem in $\mathbb{R}^2$. Ellipsoids: contour of the objective function $\boldsymbol{w}^{\dagger}\mathbf{A}\boldsymbol{w}$; Blue ellipsoid: maximum value of RQ attained when $\boldsymbol{w}=\sqrt{2}\frac{\boldsymbol{v_1}}{\left\lVert \boldsymbol{v_1}\right\rVert}$; Orange dashed circle: constraints $\boldsymbol{w}^{\dagger}\boldsymbol{w}=2$;  Cross points of an orange circle and gray dashed line: constraints in problems (\ref{10}); $\boldsymbol{v_1}$ and $\boldsymbol{v_2}$ respectively denotes the first and second eigenvalues.}
        \label{fig3}
  \end{figure}

This subsection will discuss the performance of our proposed algorithm RA, which depends on the spectrum of $\mathbf{R}$. As we know, the solution $\boldsymbol{w}_{MO}^*$ obtained by the MO can be regarded as the global optimum because the problem (\ref{first equation}) turns out to be a convex problem in terms of Riemannian geometry\cite{AbsMahSep2008}. However, it obtains the solution by searching in the manifold $\mathbb{T}^N$ so that it is hard for us to explicitly analyze the gap between $\boldsymbol{w}^{*\dagger}\mathbf{R}\boldsymbol{w}^*$ and $\boldsymbol{w}^{*\dagger}_{MO}\mathbf{R}\boldsymbol{w}^*_{MO}$. Insteadly, we will analyze the gap between $\boldsymbol{w}^{*\dagger}\mathbf{R}\boldsymbol{w}^*$ and $\boldsymbol{w}^{*\dagger}_{RQ}\mathbf{R}\boldsymbol{w}^*_{RQ}$. Due to the affiliation of the two manifolds in (\ref{affiliation}), we know that 
\begin{equation}
    \left\lvert 1-\frac{\boldsymbol{w}^{*\dagger}\mathbf{R}\boldsymbol{w}^*}{\boldsymbol{w}^{*\dagger}_{MO}\mathbf{R}\boldsymbol{w}^*_{MO}}\right\rvert \leq \left\lvert 1-\frac{\boldsymbol{w}^{*\dagger}\mathbf{R}\boldsymbol{w}^*}{\boldsymbol{w}^{*\dagger}_{RQ}\mathbf{R}\boldsymbol{w}^*_{RQ}}\right\rvert,
\end{equation}
where the performance $\alpha$ of RA is formulated as the ratio of $\boldsymbol{w}^{*\dagger}\mathbf{R}\boldsymbol{w}^*$ and $\boldsymbol{w}^{*\dagger}_{RQ}\mathbf{R}\boldsymbol{w}^*_{RQ}$. By applying the transformation $\boldsymbol{w}^*_{RQ}=\mathbf{\Sigma} \boldsymbol{w}^*$ where $\mathbf{\Sigma}=\text{diag}([a_1,\dots,a_N]^T)$ and $a_i$ is defined in Eq.(\ref{M=1}), we deifne the performance $\alpha$ as  
\begin{equation}
    \alpha
    =\frac{\lambda_1 (\boldsymbol{w}^{*\dagger}\boldsymbol{v}_1)^2+\sum_{i=2}^M \lambda_i (\boldsymbol{w}^{*\dagger}\boldsymbol{v}_i)^2}{\lambda_1 (\boldsymbol{w}^{*\dagger}_{RQ}\boldsymbol{v}_1)^2}.
    \label{performance}
\end{equation}
The projection step of the above geometry explanation would bring the value transformation from the first component to other components, which makes some performance loss compared to the RQ problem (as shown in Fig. \ref{fig3}). Usually, we believe that more perturbation of $\boldsymbol{w}^{*\dagger}_{RQ}$ results in more performance loss. So the performance of RA depends on two aspects:
\begin{itemize}
    \item the perturbation about $\boldsymbol{w}^{*\dagger}_{RQ}$, which can be measured in terms of the angle $\beta$ of $\boldsymbol{w}^*$ and $\boldsymbol{w}^{*\dagger}_{RQ}$;
    \item the scale of the first component and others, which can be measured in terms of the prominence of $\lambda_1$.
\end{itemize}

\subsubsection{The angle} 
\begin{figure}[!t]
      \centering
          {\includegraphics[width=0.8\linewidth]{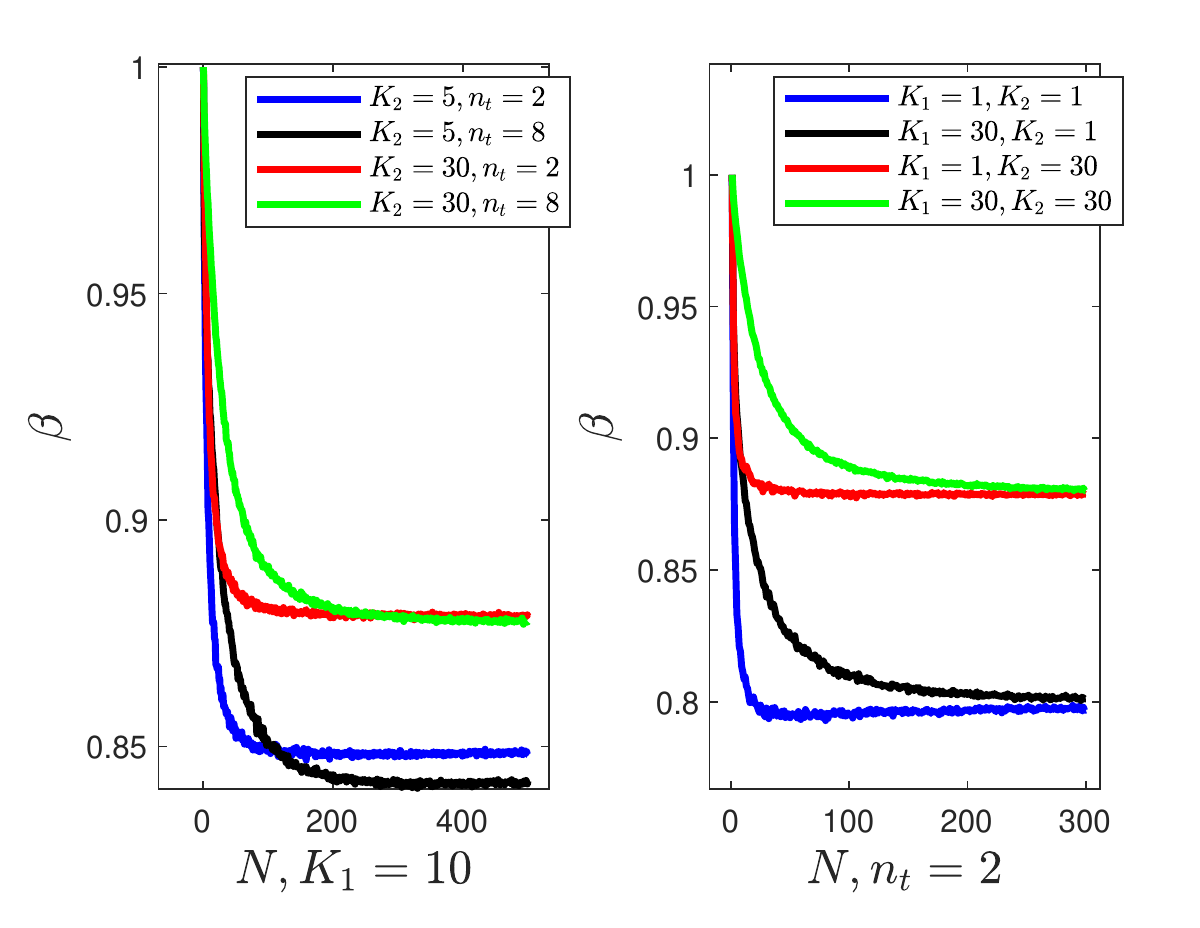}}
          \caption{Lower bound of $\beta$ obtained by numerical experiment at different $K_1$, $K_2$ and $n_t$.}
          \label{lower boundary}
\end{figure}

The angle $\beta$ of $\boldsymbol{w}^*$ and $\boldsymbol{w}^{*\dagger}_{RQ}$ can be defined as $\boldsymbol{w}^{*\dagger}\boldsymbol{v}_1/\left\lVert \boldsymbol{w}^*\right\rVert \left\lVert \boldsymbol{v}_1\right\rVert =\sum_{i=1}^N a_i/N$. We find by experiments that the mean of the distribution of $\beta$ will converge to a lower bound as $N$ increases. As shown in Fig.\ref{lower boundary}, we depict some curves to show how $n_t$, $K_1$, and $K_2$ influence $\beta$. Due to the space limitation, we refer the readers to \cite{dongxuehui} for detailed proof and discussion on this.

\textit{\textbf{Remark 2}
The lower bound of $\beta$ (defined in (\ref{performance})) increases as $K_2$ increases. The convergence rate of $\beta$ slows down as $K_1$ or $n_t$ increases. 
}


\subsubsection{The prominence of $\lambda_1$} 
Since both $\mathbf{H}_1$ and $\text{diag}(\mathbf{h}_2^T)$ are information-plus-noise model\cite{10.1214/10-AOS801}\cite{10.1214/EJP.v16-943}\cite{10.1214/009117905000000233}, $\mathbf{R}$ is a spike model which has the isolated eigenvalue.

\begin{figure}[!t]
    \centering
        {\includegraphics[width=0.86\linewidth]{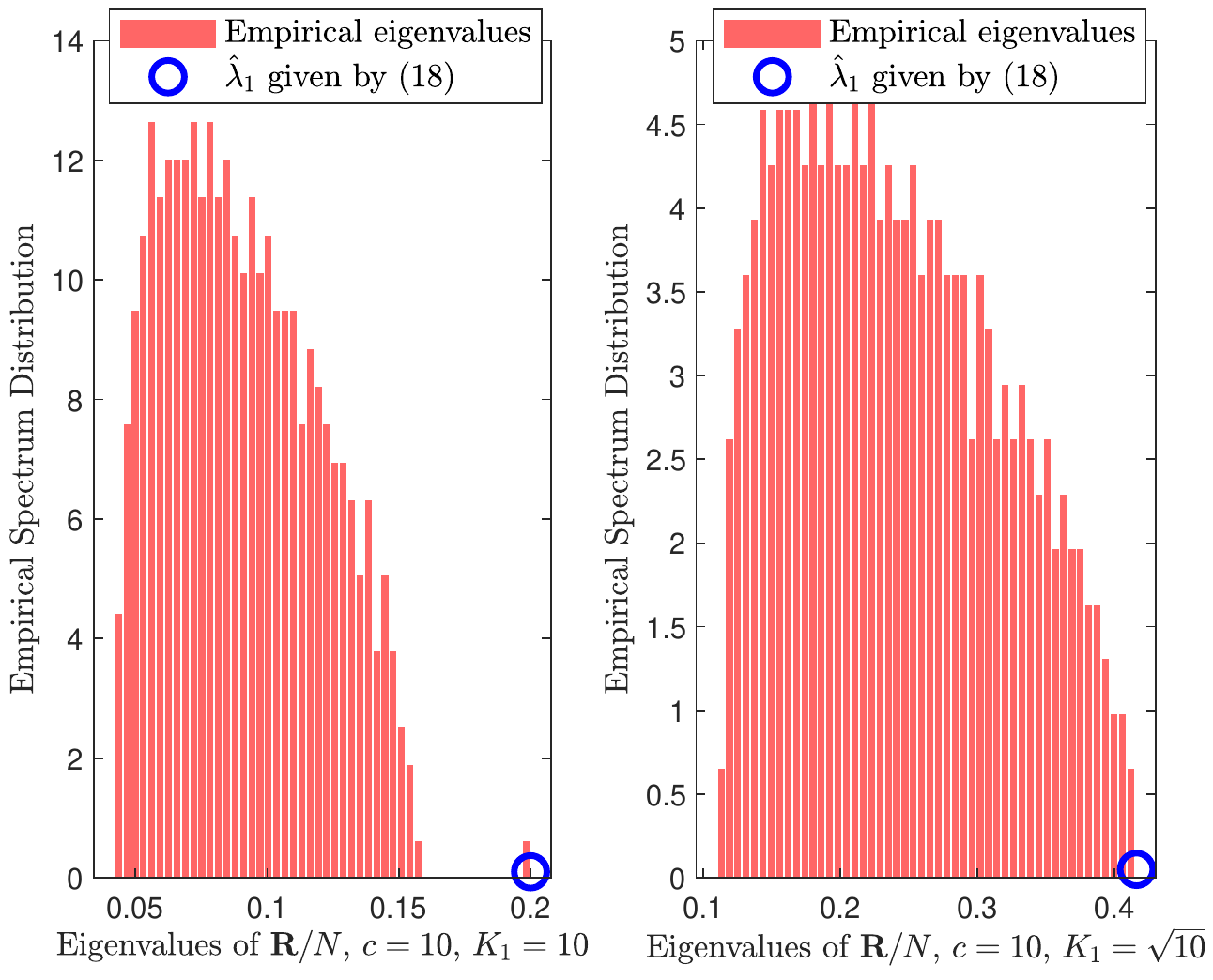}}
        \caption{The eigenvalues distributions of $\mathbf{R}$ (red histogram), the leading eigenvalue $\lambda_1$ given by Eq.(17) (blue circle).}
        \label{spike}
  \end{figure}

\textit{\textbf{Theorem 2}: 
As $n_t,N,K_2\rightarrow\infty$ such that $N/n_t\rightarrow c\in(0,\infty)$, denoting $\hat{\lambda}_1 $ the leading eigenvalues of $\mathbf{R}/N$ and the sole eigenvalue of $\mathbf{M}_1\mathbf{M}_1^{\dagger}/N$ has been defined as the number of transmitting antennas $n_t/N$ , then
\begin{equation}
    \hat{\lambda}_1 \stackrel{a . s .}{\longrightarrow} \begin{cases}\frac{1}{c}+\frac{1}{K_1} & , K_1>\sqrt{c} \\ \frac{(1+\sqrt{c})^2}{c(K_1+1)}& , K_1 \leq \sqrt{c} \end{cases}.
    \label{spike model}
\end{equation}
}
Theorem 2 identifies an abrupt change in the behavior of the prominence of the leading eigenvalues $\hat{\lambda}_1$ of $\mathbf{R}/N$ (as shown in Fig. \ref{spike}): if $K_1\leq\sqrt{c}$, where the LoS path have not dominated the channel from Tx to the RIS, the empirical spectral distribution of $\mathbf{R}$ remains unchanged meanwhile its asymptotical limit can be depicted by the Mar\v{c}enko-Pastur distribution $\mu$\cite{couillet_liao_2022}. However, as soon as $K_1>\sqrt{c}$, $\hat{\lambda}_1$ converges to a limit $\frac{1}{c}+\frac{1}{K_1}$ beyond the right-edge $\frac{(1+\sqrt{c})^2}{c(K_1+1)}$ of $\mu$, and the leading eigenvalue becomes more prominent as $K_1$ increasing. In addition, as mentioned in Theorem 1, the performance of RA is exactly the global optimum when $K_1\rightarrow\infty$ (i.e., $\mathbf{R}$ becomes a rank-one matrix).

\textit{\textbf{Remark 3}: The max SNR obtained by the RA will approach the upper bound of it (obtained by the MO) as the LoS path dominates the channel (i.e., $K_1\rightarrow\infty$)
}

  \section{Validation of the Relaxation Algorithm via Numerical Simulation}
  In this section, numerical results are provided to validate the proposed RA's effectiveness. We will show the extremely low time consumption and high reliability for high-rank $\mathbf{R}$ through simulation. The simulation is performed in a Matlab environment in Windows 11 operating system with CPU i7-12700k, and the code has been open-sourced to the GitHub website\footnote{https://github.com/DwyaneDong/relaxation-algorithm-on-RIS-MISO.git}.
  
  Note that the time consumed of SDR is linearly correlated with the number of Gaussian random vectors we generated\cite{8647620} (we generated 25 gaussian random vectors every loop in Fig. 6 experiment setting and only one in Fig. 7 settings). Moreover, the SNR obtained by MO can be regarded as the upper bound.
  \begin{figure}[htp]
      \centering
      {\includegraphics[width=0.75\linewidth]{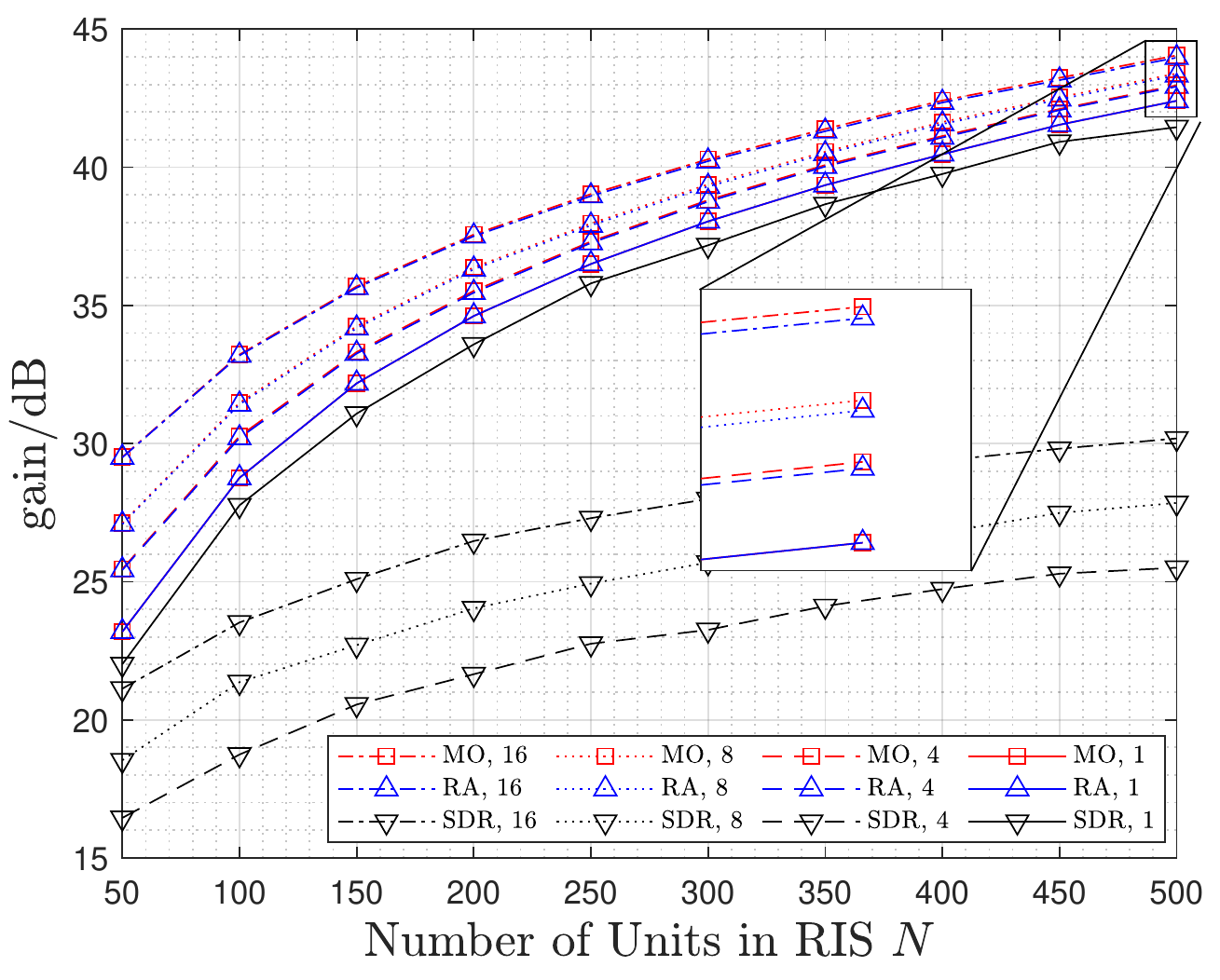}}
      \caption{The SNR obtained by RA, MO, and SDR in four different ranks (1, 4, 8, 16) of $\mathbf{R}$ while the number of units grows from 50 to 500. }
      \label{SNR_fig}
  \end{figure}
  
  The SNR we will evaluate is defined in (\ref{SNR}). The Fig.\ref{SNR_fig} shows the SNRs of three methods for different ranks of $\mathbf{R}$ whose dimensions grow from 50 to 500, averaged over $10^3$ channel realization. From this figure, the proposed RA method obtains exactly the same SNR as the MO method while rank is one, which proves Theorem 1 that the solution given by (\ref{close-form}) is the global optimum for the rank-one situation. And RA methods can reach respectively $100\%$, $99.1\%, 98.4\%$ and $98.2\%$ of the SNRs obtained by the MO methods in rank 1, 4, 8, and 16 when $N$ is 500. These results prove Theorem 1 in experiments where we find the closed-form global optimal solution of the rank-one situation. Even though we increased the number of generated Gaussian vectors to over $10^2$ in simulation\cite{8647620}, the SDR method has difficulty in achieving SNR as good as the first two methods, and the higher the rank, the larger the difference in SNR. The reason why SDR failed has been discussed in\cite{8647620}.

\begin{figure}[htp]
      \centering
      {\includegraphics[width=0.75\linewidth]{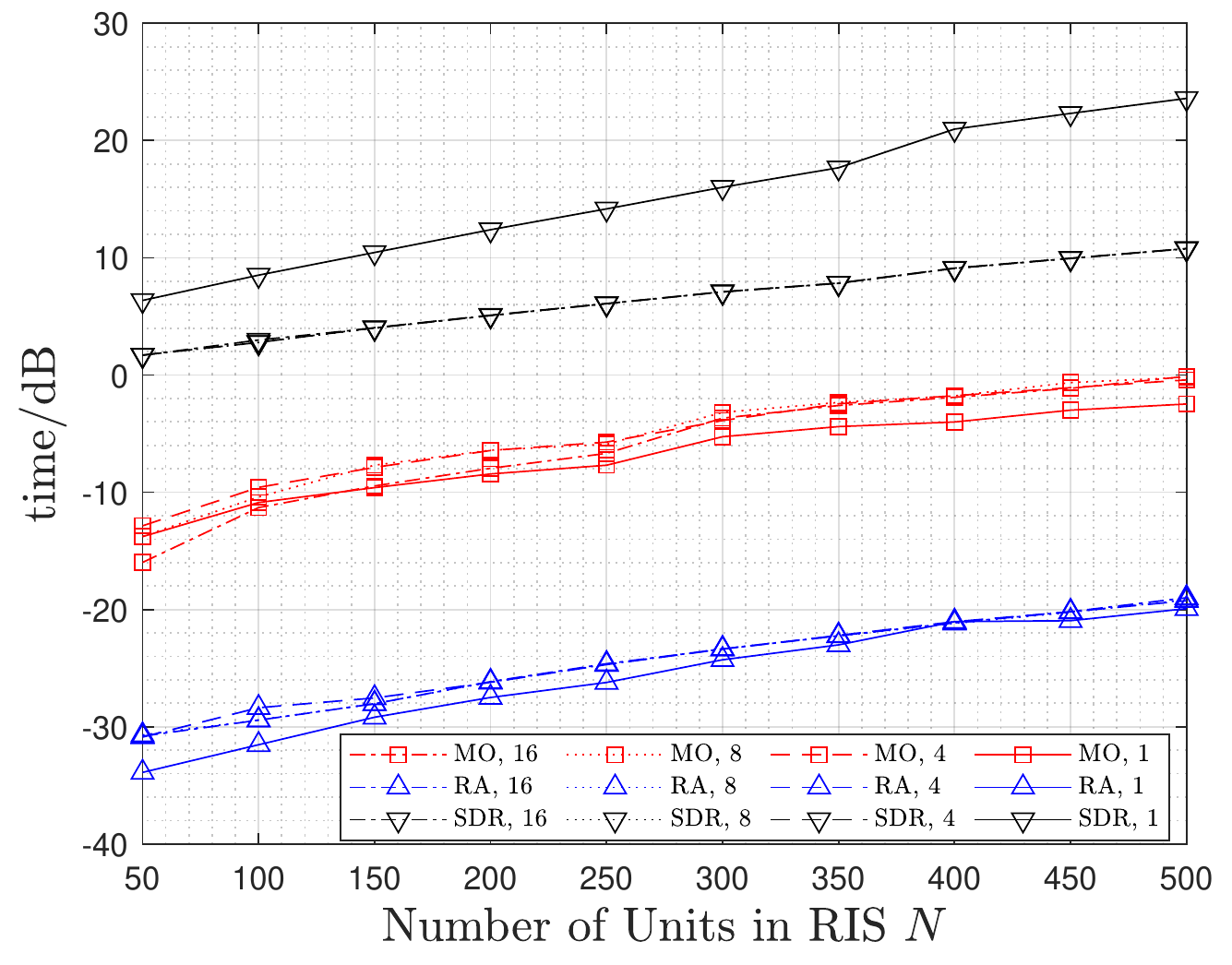}}
      \caption{The time consumption of the RA and the MO in four different ranks (1, 4, 8, 16) of $\mathbf{R}$ while the number of units grows from 50 to 500}
      \label{time}
  \end{figure}

  The most interesting part about RA is the extremely low time consumption which comes from its simplicity. We perform this comparison using the open source toolboxes Manopt\cite{DBLP:journals/corr/BoumalMAS13} and CVX\cite{cvx}. As shown in Fig. \ref{time}, the time consumption of RA is below $1\%$ of that of MO and below $0.1\%$ of that of SDR. Unlike all the other algorithms, the time consumption of RA does not depend on the rank of $\mathbf{R}$. This exciting property makes it possible to perform passive beamforming for RIS-assisted communication in a high-mobility channel environment where the coherence time is very short.
  
\begin{figure}[htp]
      \centering
      {\includegraphics[width=0.75\linewidth]{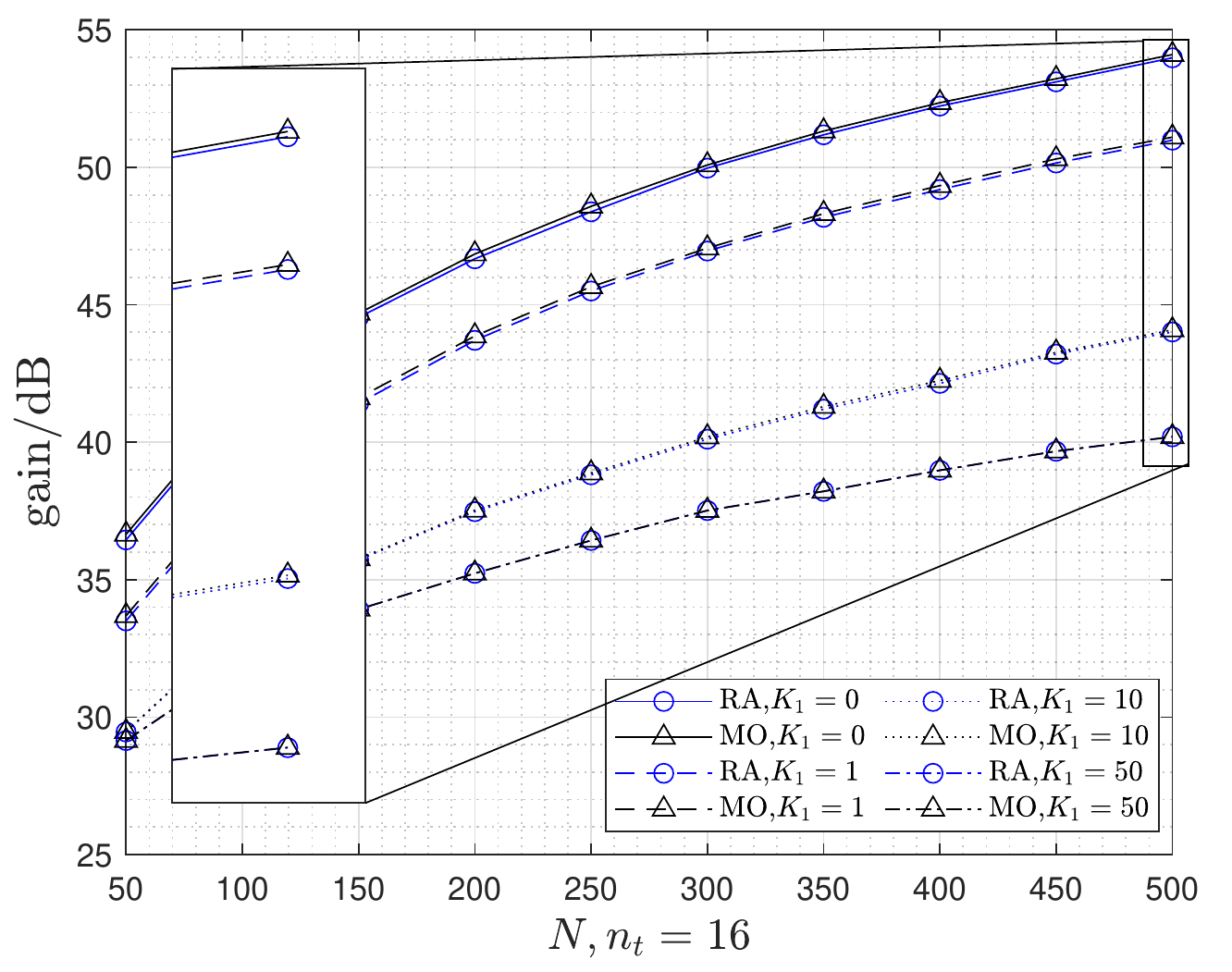}}
      \caption{The performance of the RA increases as the $K_1$ increasing and remains unchanged as $N$ increasing.}
      \label{RA_2}
  \end{figure}

  As shown in Fig. \ref{RA_2}, the performance of RA will approach the global optimum (obtained by the MO) as $K_1$ increasing. The RA-MO SNR ratio is respectively $97.30\%$, $97.50\%$, $98.33\%$ and $99.87\%$ as the $K_1$ equals to $0$, $1$, $10$ and $50$. And the ratio is hardly changed by the number of units in RIS when $n_t$ and $K_1$ stay the same.
  
\begin{figure}[htp]
    \centering
    {\includegraphics[width=0.75\linewidth]{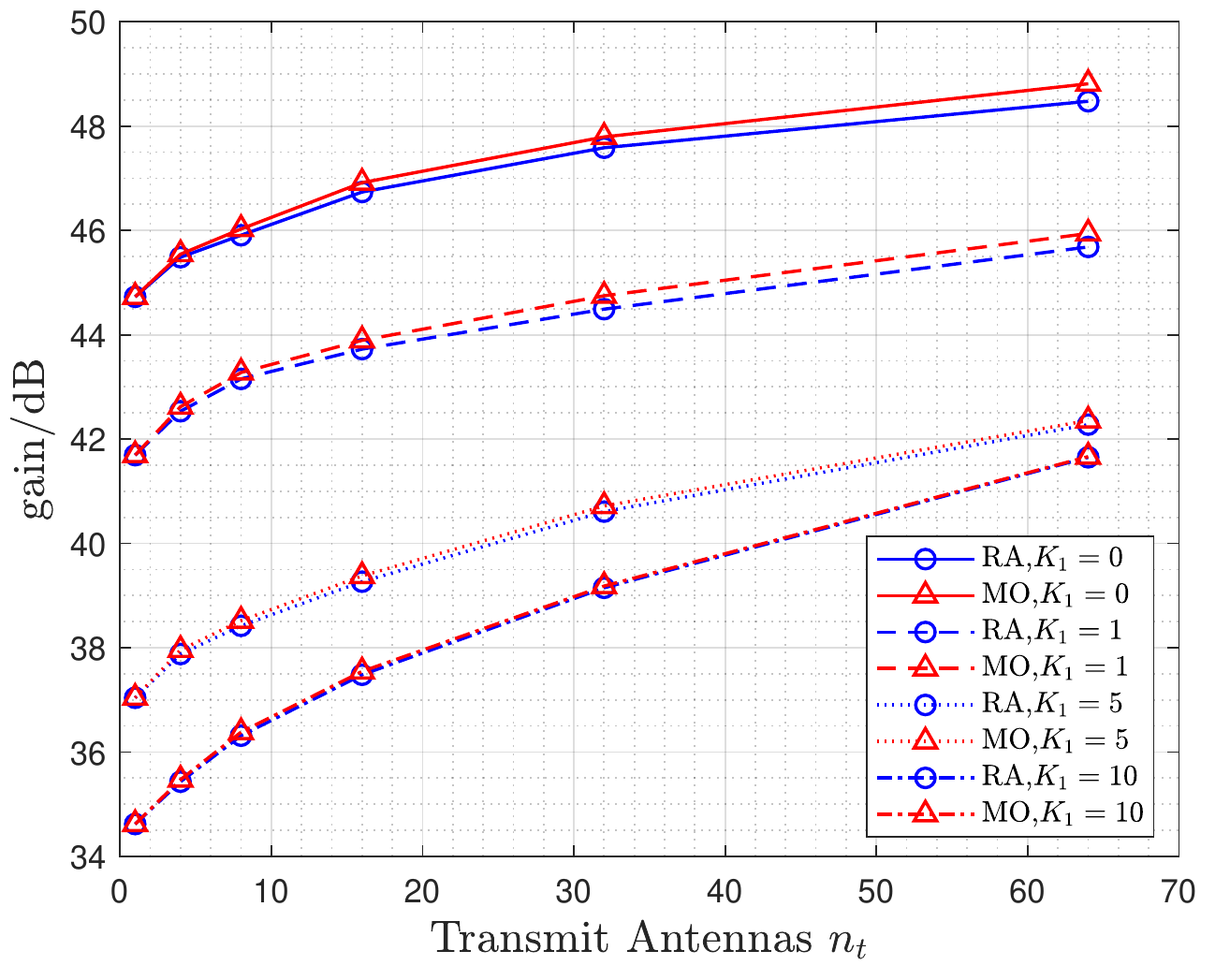}}
    \caption{The performance of the RA increases as the $K_1$ increasing and slowly decreases as $n_t$ increasing.}
    \label{RA_4}
    \end{figure}

  The Fig. \ref{RA_4} shows that the performance slowly decreases compared to the global optimum (obtained by the MO). At the same time, the number $n_t-1$ of non-leading and non-zero eigenvalues of $\mathbf{R}$ increases, the sum of spectral components except the leading one will increase because the norm of the projection of $\boldsymbol{w}^*$ onto any other eigenvector converges to a constant.

\section{conclution and perspective}
In this paper, we proved that the rank-one passive beamforming for RIS-assisted communication has a closed-form optimal solution. We proposed a relaxation algorithm (i.e., RA) that consumes an extremely short time to get a great performance. The RA truly takes the passive beamforming for RIS-assisted communication systems from tedious to simple. 
\bibliographystyle{IEEEtran}
\bibliography{ref_paper}

\begin{thebibliography}{10}
\providecommand{\url}[1]{#1}
\csname url@samestyle\endcsname
\providecommand{\newblock}{\relax}
\providecommand{\bibinfo}[2]{#2}
\providecommand{\BIBentrySTDinterwordspacing}{\spaceskip=0pt\relax}
\providecommand{\BIBentryALTinterwordstretchfactor}{4}
\providecommand{\BIBentryALTinterwordspacing}{\spaceskip=\fontdimen2\font plus
\BIBentryALTinterwordstretchfactor\fontdimen3\font minus
  \fontdimen4\font\relax}
\providecommand{\BIBforeignlanguage}[2]{{%
\expandafter\ifx\csname l@#1\endcsname\relax
\typeout{** WARNING: IEEEtran.bst: No hyphenation pattern has been}%
\typeout{** loaded for the language `#1'. Using the pattern for}%
\typeout{** the default language instead.}%
\else
\language=\csname l@#1\endcsname
\fi
#2}}
\providecommand{\BIBdecl}{\relax}
\BIBdecl

\bibitem{dang2020should}
S.~Dang, O.~Amin, B.~Shihada, and M.-S. Alouini, ``What should 6g be?''
  \emph{Nature Electronics}, vol.~3, no.~1, pp. 20--29, 2020.

\bibitem{9136592}
C.~Huang, S.~Hu, G.~C. Alexandropoulos, A.~Zappone, C.~Yuen, R.~Zhang, M.~D.
  Renzo, and M.~Debbah, ``Holographic mimo surfaces for 6g wireless networks:
  Opportunities, challenges, and trends,'' \emph{IEEE Wireless Communications},
  vol.~27, no.~5, pp. 118--125, 2020.

\bibitem{9133142}
C.~You, B.~Zheng, and R.~Zhang, ``Channel estimation and passive beamforming
  for intelligent reflecting surface: Discrete phase shift and progressive
  refinement,'' \emph{IEEE Journal on Selected Areas in Communications},
  vol.~38, no.~11, pp. 2604--2620, 2020.

\bibitem{9551980}
X.~Pei, H.~Yin, L.~Tan, L.~Cao, Z.~Li, K.~Wang, K.~Zhang, and E.~Björnson,
  ``Ris-aided wireless communications: Prototyping, adaptive beamforming, and
  indoor/outdoor field trials,'' \emph{IEEE Transactions on Communications},
  vol.~69, no.~12, pp. 8627--8640, 2021.

\bibitem{8647620}
Q.~Wu and R.~Zhang, ``Intelligent reflecting surface enhanced wireless network:
  Joint active and passive beamforming design,'' in \emph{2018 IEEE Global
  Communications Conference (GLOBECOM)}, 2018, pp. 1--6.

\bibitem{8741198}
C.~Huang, A.~Zappone, G.~C. Alexandropoulos, M.~Debbah, and C.~Yuen,
  ``Reconfigurable intelligent surfaces for energy efficiency in wireless
  communication,'' \emph{IEEE Transactions on Wireless Communications},
  vol.~18, no.~8, pp. 4157--4170, 2019.

\bibitem{9779545}
Y.~Zhang, K.~Shen, S.~Ren, X.~Li, X.~Chen, and Z.-Q. Luo, ``Configuring
  intelligent reflecting surface with performance guarantees: Optimal
  beamforming,'' \emph{IEEE Journal of Selected Topics in Signal Processing},
  vol.~16, no.~5, pp. 967--979, 2022.

\bibitem{9110869}
C.~Huang, R.~Mo, and C.~Yuen, ``Reconfigurable intelligent surface assisted
  multiuser miso systems exploiting deep reinforcement learning,'' \emph{IEEE
  Journal on Selected Areas in Communications}, vol.~38, no.~8, pp. 1839--1850,
  2020.

\bibitem{8855810}
X.~Yu, D.~Xu, and R.~Schober, ``Miso wireless communication systems via
  intelligent reflecting surfaces : (invited paper),'' in \emph{2019 IEEE/CIC
  International Conference on Communications in China (ICCC)}, 2019, pp.
  735--740.

\bibitem{9405423}
M.~A. ElMossallamy, K.~G. Seddik, W.~Chen, L.~Wang, G.~Y. Li, and Z.~Han, ``Ris
  optimization on the complex circle manifold for interference mitigation in
  interference channels,'' \emph{IEEE Transactions on Vehicular Technology},
  vol.~70, no.~6, pp. 6184--6189, 2021.

\bibitem{AbsMahSep2008}
P.-A. Absil, R.~Mahony, and R.~Sepulchre, \emph{Optimization Algorithms on
  Matrix Manifolds}.\hskip 1em plus 0.5em minus 0.4em\relax Princeton, NJ:
  Princeton University Press, 2008.

\bibitem{8811733}
Q.~Wu and R.~Zhang, ``Intelligent reflecting surface enhanced wireless network
  via joint active and passive beamforming,'' \emph{IEEE Transactions on
  Wireless Communications}, vol.~18, no.~11, pp. 5394--5409, 2019.

\bibitem{5447068}
Z.-q. Luo, W.-k. Ma, A.~M.-c. So, Y.~Ye, and S.~Zhang, ``Semidefinite
  relaxation of quadratic optimization problems,'' \emph{IEEE Signal Processing
  Magazine}, vol.~27, no.~3, pp. 20--34, 2010.

\bibitem{dongxuehui}
X.~Dong, ``Spectral analysis of relaxation algorithm for unit circle
  constrained complex quadratic form,'' \emph{in preparation}, 2023.

\bibitem{10.1214/10-AOS801}
\BIBentryALTinterwordspacing
N.~E. Karoui, ``{On information plus noise kernel random matrices},'' \emph{The
  Annals of Statistics}, vol.~38, no.~5, pp. 3191 -- 3216, 2010. [Online].
  Available: \url{https://doi.org/10.1214/10-AOS801}
\BIBentrySTDinterwordspacing

\bibitem{10.1214/EJP.v16-943}
\BIBentryALTinterwordspacing
P.~Loubaton and P.~Vallet, ``{Almost Sure Localization of the Eigenvalues in a
  Gaussian Information Plus Noise Model. Application to the Spiked Models.}''
  \emph{Electronic Journal of Probability}, vol.~16, no. none, pp. 1934 --
  1959, 2011. [Online]. Available: \url{https://doi.org/10.1214/EJP.v16-943}
\BIBentrySTDinterwordspacing

\bibitem{10.1214/009117905000000233}
\BIBentryALTinterwordspacing
J.~Baik, G.~B. Arous, and S.~P{\'e}ch{\'e}, ``{Phase transition of the largest
  eigenvalue for nonnull complex sample covariance matrices},'' \emph{The
  Annals of Probability}, vol.~33, no.~5, pp. 1643 -- 1697, 2005. [Online].
  Available: \url{https://doi.org/10.1214/009117905000000233}
\BIBentrySTDinterwordspacing

\bibitem{couillet_liao_2022}
R.~Couillet and Z.~Liao, \emph{Random Matrix Methods for Machine
  Learning}.\hskip 1em plus 0.5em minus 0.4em\relax Cambridge University Press,
  2022.

\bibitem{DBLP:journals/corr/BoumalMAS13}
\BIBentryALTinterwordspacing
N.~Boumal, B.~Mishra, P.~Absil, and R.~Sepulchre, ``Manopt, a matlab toolbox
  for optimization on manifolds,'' \emph{CoRR}, vol. abs/1308.5200, 2013.
  [Online]. Available: \url{http://arxiv.org/abs/1308.5200}
\BIBentrySTDinterwordspacing

\bibitem{cvx}
M.~Grant and S.~Boyd, ``{CVX}: Matlab software for disciplined convex
  programming, version 2.1,'' \url{http://cvxr.com/cvx}, Mar. 2014.

\end{thebibliography}
\vspace{12pt}
\end{document}